\documentclass[fleqn,usenatbib]{mnras}

\usepackage[T1]{fontenc}

\DeclareRobustCommand{\VAN}[3]{#2}
\let\VANthebibliography\thebibliography
\def\thebibliography{\DeclareRobustCommand{\VAN}[3]{##3}\VANthebibliography}

\usepackage{array,multirow,graphicx} % Including figure files
\usepackage{float} 
\usepackage{amsmath}  % Advanced maths commands
\usepackage{amssymb}  % Extra maths symbols
\usepackage{physics}
\usepackage{tabularx}
\usepackage{xcolor}
\usepackage{tikz}

\definecolor{lime}{HTML}{A6CE39}
\DeclareRobustCommand{\orcidicon}{%
  \begin{tikzpicture}
  \draw[lime, fill=lime] (0,0) 
  circle [radius=0.16] 
  node[white] {{\fontfamily{qag}\selectfont \tiny ID}};
  \draw[white, fill=white] (-0.0625,0.095) 
  circle [radius=0.007];
  \end{tikzpicture}
  \hspace{-2mm}
}

\foreach \x in {A, ..., Z}{%
  \expandafter\xdef\csname orcid\x\endcsname{\noexpand\href{https://orcid.org/\csname orcidauthor\x\endcsname}{\noexpand\orcidicon}}
}

\usepackage{newtxtext,newtxmath}
\title[Gravitational lensing modification of the UVLF]{Gravitational lensing modification of the high redshift galaxy luminosity function}

\author[G. Ferrami \& J.S.B. Wyithe]{
G. Ferrami $^{1,}$$^{2}$ \thanks{E-mail: gferrami@student.unimelb.edu.au } \orcidA{}
J. Stuart B. Wyithe $^{1,}$$^{2}$ \orcidB{}\\
$^{1}$School of Physics, University of Melbourne, Parkville, VIC 3010, Australia\\
$^{2}$ARC Centre of Excellence for All-Sky Astrophysics in 3 Dimensions (ASTRO 3D)
}

\date{Accepted XXX. Received YYY; in original form ZZZ}

\pubyear{2022}

\begin{document}
\label{firstpage}
\pagerange{\pageref{firstpage}--\pageref{lastpage}}
\maketitle

% Abstract of the paper - 250 words (200 words for Letters)
\begin{abstract}
The bright end of the rest-frame UV luminosity function (UVLF) of high-redshift galaxies is modified by gravitational lensing magnification bias. 
Motivated by recent discoveries of very high-z galaxies with JWST, we study the dependence of magnification bias on the finite size of sources at $6<z<14$.
We calculate the magnification probability distributions and use these to calculate the magnification bias assuming a rest-frame Schechter UVLF for galaxies at redshift $6<z<14$.
We find that the finite size of bright high-redshift galaxies together with lens ellipticity significantly suppresses magnification bias, producing an observed bright end which declines more sharply than the power-law resulting from assumption of point sources. 
By assuming a luminosity-size relation for the source population and comparing with the observed $z=6$ galaxy luminosity function from \cite{Harikane_2022}, we show that the UVLF can be used to set mild constraints on the galaxies intrinsic size, favoring smaller galaxies compared to the fiducial luminosity-size relation. In the future, wide surveys using \textit{Euclid} and \textit{Roman Space Telescope} will place stronger constraints.
We also tabulate the maximum magnification possible as a function of source size and lens ellipticity.
\end{abstract}


%%%%%%%%%%%%%%%%%%%%%%%%%%%%%%%%%%%%%%%%%%%%%%%%%%

%%%%%%%%%%%%%%%%% BODY OF PAPER %%%%%%%%%%%%%%%%%%

\section{Introduction}

Study of the rest-frame UV luminosity function (UVLF) is fundamental to understanding the role of high-redshift galaxies during the epoch of reionization and their evolution through cosmic time.
In recent years, surveys have measured the UVLF over a large absolute magnitude interval up to $z\approx 9$ (e.g. \citealt{Bouwens_2022},  \citealt{Harikane_2022}, \citealt{Atek_2018}, \citealt{Livermore_2017}, \citealt{Finkelstein_2015}).
These works suggest that the rest-frame UVLF is well fitted by a Schechter function (a power law at faint luminosities and an exponential cut-off at bright luminosities, \citealt{Schechter_1976}) or a double power-law, whose parameters evolve with redshift.\\
\indent The number counts and flux measurements of high-redshift sources are distorted by gravitational lensing due to foreground structures (\citealt{Barkana_Loeb_2000}, \citealt{Wyithe_Nature_2011}, \citealt{Mason_2015}).
In particular, considering a Schechter profile for the intrinsic UVLF, the magnification of more numerous fainter galaxies leads to an apparent increase in the observed number of bright galaxies, an effect called magnification bias (\citealt{Turner_1984}).
Other studies focused on modeling the low magnification regime for virialized and non-virialized lens structures (\citealt{Fialkov_2015}).
\cite{Barone_Nugent_2015} directly measured the magnification bias on a sample of Lyman-break galaxies at $4\leq z\leq 8$ in the eXtreme Deep Field (XDF) and the Cosmic Assembly Near-infrared Deep Extragalactic Legacy Survey (CANDELS) fields.\\
\indent Modification of the statistics of the bright end of the observed high-redshift galaxy UVLF are dominated by galaxy-scale lenses, which are well described by isothermal ellipsoids (\citealt{Keeton_1997}). Background galaxy sources are finite-sized with a luminosity profile characterized by a characteristic radius.
Both non-spherical lenses and extended sources affect the predictions of a lensing model (e.g. see \citealt{Schneider_Ehlers_Falco_book} for a theoretical introduction and \citealt{Oguri_2002} for an application on the statistics of strong lensing radial arcs). 
However, discussion of lensing statistics for high-z galaxies has been restricted to spherical lenses and point sources.\\
\indent In this \textit{Letter} we study the combined effect of elliptical lenses and extended sources on the bright end of the luminosity function for high-redshift galaxies.
In Section 2, we obtain magnification probability distributions for extended sources assuming a singular isothermal ellipsoid (SIE) lens model.
In Section 3, we calculate apparent luminosity functions and the magnification bias assuming a rest-frame Schechter UVLF for galaxies at redshift $6<z<14$. 
We then compare our models with some previous observational results for the UV luminosity functions for galaxies at $z=6$ and show how this could lead to constraints on the intrinsic size distribution for high-redshift galaxies.
Discussion and conclusions are presented in Section 4.
Throughout this paper, we adopt $H_0 = 70$ km s$^{-1}$ Mpc$^{-1}$, $\Omega_0 = 0.3$,  $\Omega_\Lambda= 0.7$.

\begin{figure*}
  \includegraphics[width=0.9\linewidth]{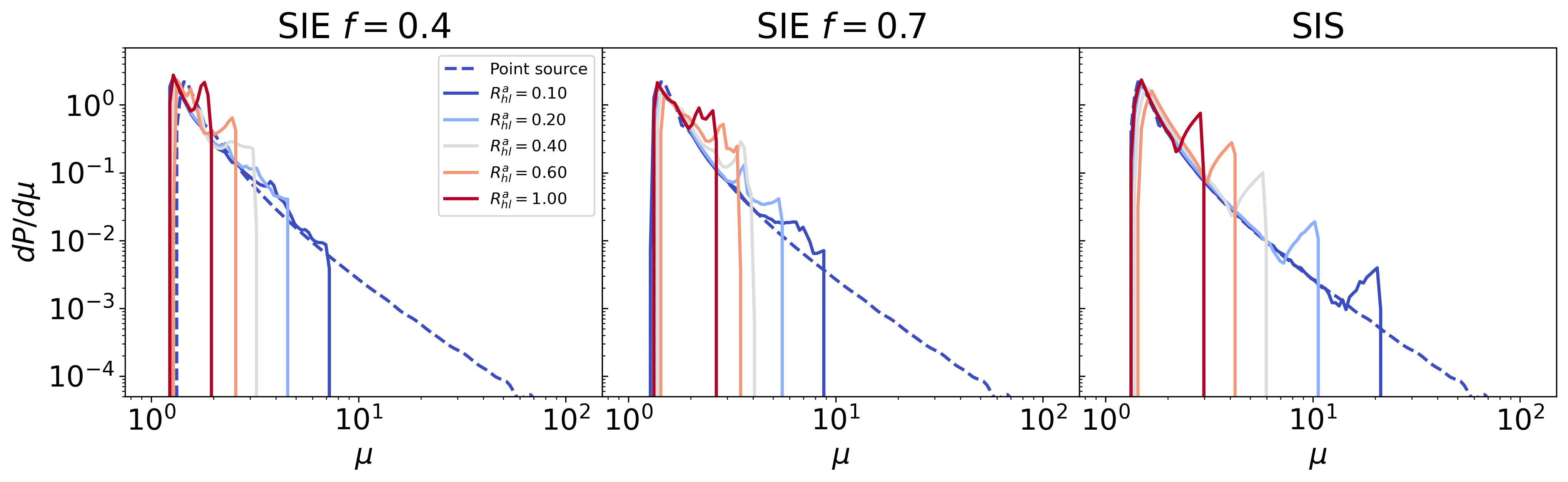}
    \caption{Magnification probability distribution for the brightest image produced by SIE lenses with axis ratios $f=0.3$ (left), $f=0.7$ (center) and $f=1$ (SIS model, right). 
    The magnification probability distributions for point-like sources share the $\dv{P}{\mu}\sim \mu^{-3}$ trend characteristic of the SIS profile (e.g. see \citealt{Wyithe_Nature_2011}), while the magnification probability distributions for extended sources show a cutoff after a certain maximum value of magnification inversely related
     to the dimensionless half-light radius of the sources $R_{hl}^{a}$.}
    \label{fig:dPdMu_ellipticity}
\end{figure*}

\section{Magnification probability distributions}

The dimensionless lens equation relating source position $\vb*y$ to image positions $\vb*x$ through the deflection angle $\vb*\alpha(\vb*x)$ is
\begin{equation}\label{eq:lens_equation_dimensionless}
  \vb*y = \vb*x - \vb*\alpha(\vb*x) \text{ ,}
\end{equation}
where $\vb*x$ and $\vb*y$ are expressed in units of the Einstein radius, which, for a singular isothermal sphere lens model (SIS, see for example \citealt{Schneider_Ehlers_Falco_book}), is
\begin{equation}\label{eq:Einstein_radius}
  \theta_E(\sigma,z) = 4\pi\frac{D_{ls}}{D_s}\left(\frac{\sigma}{c}\right)^2 = 1^{\prime\prime} \frac{D_{ls}}{D_s}\left(\frac{\sigma}{186 \text{ km/s}}\right)^2 \text{ .}
\end{equation}
Here $\sigma$ is the velocity dispersion. Therefore $\theta_E\sim 1^{\prime\prime}$ in galaxy-scale lenses with a geometry such that ${D_{ls}}/{D_s}\approx 1$.
The cosmological distances $D_l$, $D_s$, $D_{ls}$ are respectively the angular diameter distances between the observer and the deflector, the observer and the source, and the deflector and the source.\\
\indent To calculate the magnification, we consider a singular isothermal ellipsoid (SIE, see \citealt{Kormann_Schneider_Bartelmann_1994}) model for the lens, characterized by the parameter $f$ defined as the ratio between the minor and major axes of the elliptical isodensity contours\footnote{The other model parameters are ignored here, since the connection with the value of the surface mass density is removed by the dimensionless scaling, and our analysis is not sensible to the absolute position and rotation of the lens model.}. 
Due to the singularity of the surface mass density, a SIE lens produces 1, 2 or 4 images of a point source, depending on the position of the source with respect to lens.\\
\indent Gravitational lensing conserves surface-brightness $I$ but stretches the solid angle that an unlensed source subtends in the sky. The ratio of the observed and intrinsic (unlensed) fluxes gives the magnification $\mu$. 
For an extended source, one integrates the point-like magnification over the surface-brightness profile of the source $I(\vb* y)$ to obtain 

\begin{equation}\label{eq:mag_ext_general}
  \mu_e=\frac{\int_{R^2}\dd ^2y I(\vb*y) \mu_p(\vb*y)}{\int_{R^2}\dd ^2y I(\vb* y)}\text{ .}
\end{equation}
\indent An extended source therefore has a lower or equal magnification than a point-source through the same lens.
For example, considering a circular source with a constant surface-brightness near a linear caustic, it has been shown that there is a maximum magnification at a finite distance from the caustic (see Appendix of \citealt{Schneider_1987}).
This implies that the magnification probability distribution has a cutoff after $\mu_{max}$. \\
\indent Our choice of modeling elliptical lenses and extended sources improves upon the previous statistical lensing studies on the luminosity function of high-redshift galaxies, which considered point-like sources and a singular isothermal sphere model for the lenses (\citealt{Wyithe_Nature_2011}, \citealt{Mason_2015}).
The effect of the finite size of a source becomes significant when it is comparable with its distance from the caustic.
We expect bright galaxies at high-redshift to have angular sizes $\lesssim 0.5^{\prime \prime}$ (e.g. see \citealt{Liu_Wyithe_Dragons_VII_2017} for a semi-analytical model of high-z galaxies size), which is smaller than but comparable to the Einstein radius $\theta_E$.\\
\indent We implemented the numerical solution for the SIE lens model discussed in \cite{Kormann_Schneider_Bartelmann_1994}, initializing a dimensionless source plane with a $l^2$ square area divided in $1000\times1000$ pixels.
We chose the value to be $l=4.2$, to have both high resolution in the high magnification regions and explore the low magnification regions away from the caustic.
We solve the lens equation assuming different choices of the lens axis ratio $f$ for a point-like source in each pixel, and retrieved a map of the magnification values for the brightest of the images of each source.
We then applied Eq. (\ref{eq:mag_ext_general}) considering a circular source with a constant surface-brightness and different choices for the values of the  dimensionless half-light radius $R^{a}_{hl}$ of the source, obtained by scaling its physical size $R_{hl}$ in the source plane as $R_{hl}^{a} = \frac{R_{hl}}{\theta_E} \frac{D_s}{D_l}$.
From these magnification maps, we obtained the magnification probability distributions for the brightest image as a function of $f$ and $R^{a}_{hl}$. \\
\indent Some of those are represented in Fig. \ref{fig:dPdMu_ellipticity}. 
The magnification probability distributions for point-like sources show the expected power-law characteristic of the SIS profile (e.g. see \citealt{1995_Pei}, \citealt{Wyithe_Nature_2011}), while the magnification probability distributions for extended sources present a cutoff above a certain maximum value of magnification. This maximum increases toward small sources and spherical lenses, and will be of interest in particular observed cases.
The values of maximum magnification as a function of source size and lens ellipticity are listed in Table \ref{tab:max_magnification}.

\begin{table}
\caption{Maximum magnification for an extended source.}
\label{tab:max_magnification}
\begin{tabular}{|p{0.1cm}|p{0.3cm}||p{0.45cm}|p{0.45cm}|p{0.45cm}|p{0.45cm}|p{0.45cm}|p{0.45cm}|p{0.45cm}|p{0.45cm}|}
\cline{3-10}
\multicolumn{2}{c|}{}&\multicolumn{8}{c|}{Source dimensions [arcsec $\times (1"/\theta_E) \times (D_s/D_l)$]}\\
\cline{3-10}
\multicolumn{2}{c|}{}&0.05&0.10&0.15&0.20&0.40&0.60&0.80&1.00\\
\hline
\hline
\parbox[t]{2mm}{\multirow{11}{*}{\rotatebox[origin=c]{90}{Lens axes ratio $f\times10$}}} 
&1&5.74&3.62&2.87&2.46&1.95&1.49&1.23&1.18\\
\cline{2-10}
&2&9.12&5.97&4.74&4.06&2.66&2.03&1.68&1.49\\
\cline{2-10}
&3&10.23&6.70&5.32&4.39&2.98&2.46&2.11&1.81\\
\cline{2-10}
&4&11.05&7.24&5.53&4.56&3.22&2.56&2.28&1.95\\
\cline{2-10}
&5&12.41&7.82&5.97&4.92&3.35&2.66&2.28&2.19\\
\cline{2-10}
&6&13.40&8.44&6.45&5.74&3.48&2.87&2.66&2.46\\
\cline{2-10}
&7&14.47&8.77&7.52&5.53&4.06&3.48&2.98&2.66\\
\cline{2-10}
&8&16.24&11.49&7.52&6.20&4.92&3.91&3.22&2.76\\
\cline{2-10}
&9&22.97&12.89&11.94&9.85&5.74&4.22&3.35&2.87\\
\cline{2-10}
&SIS&42.54&21.27&14.47&10.64&5.97&4.22&3.48&2.98\\
\hline
\end{tabular}
\end{table}

\section{Lensed luminosity functions of extended sources}
\begin{figure*}
  \includegraphics[width=0.85\linewidth]{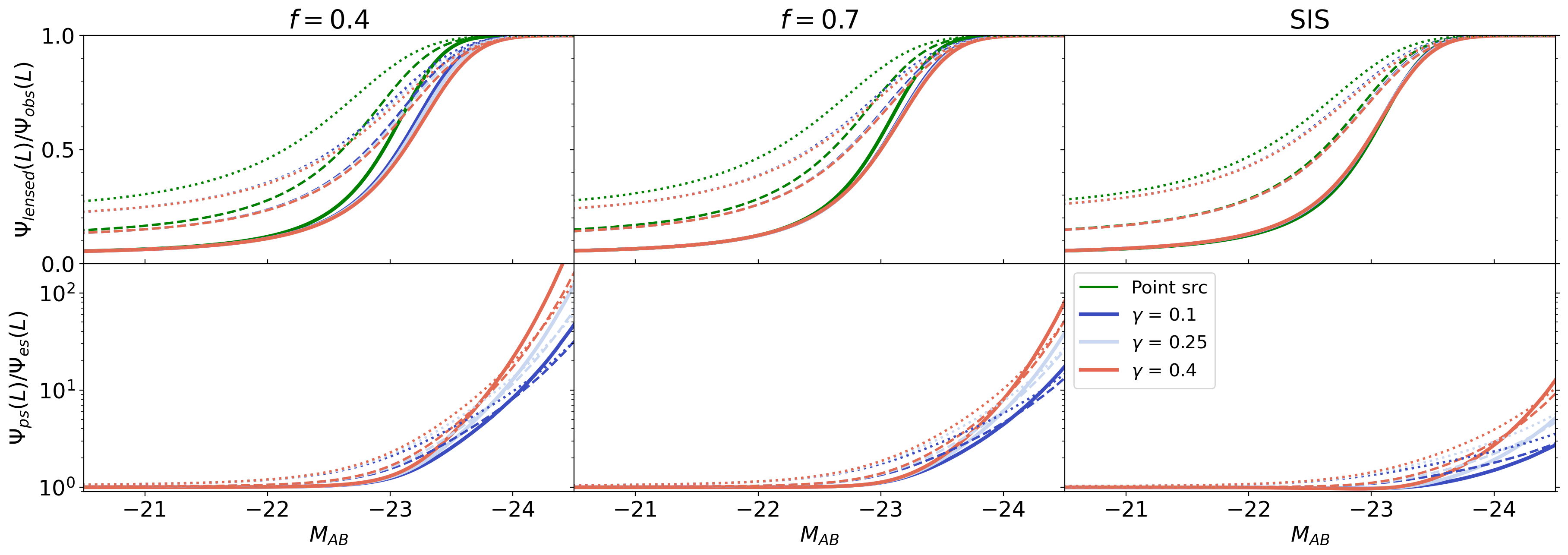}
    \caption{(Upper) Ratio between the lensed UVLF and observed UVLF $\Psi_{lensed}/\Psi_{obs}$, which in the bright end approximates the fraction of observed lensed galaxies per luminosity interval for different L-size relations at $z=7$ (solid, thin dashed and dotted lines are for redshift $z=11$ and $z=14$).
    (Lower) Ratio between the apparent UVLF for a point-source and the apparent UVLF for extended sources with different choices for the value of $\gamma$ in the L-size relation, with $R_0=0.11"$.
    On the left side we show the results assuming a population of SIE lenses with axis ratio $f=0.4$, in the center with $f=0.7$, and on the right assuming a population of spherical lenses.}
    \label{fig:mag_bias_and_ratio}
\end{figure*}

The \textit{a-priori} probability for a source to be lensed by a foreground galaxy into multiple images is defined as the optical depth $\tau_m$ (e.g., see \citealt{Wyithe_Nature_2011} and references therein), defined as
\begin{equation}\label{eq:optical_depth}
\begin{split}
  \tau_m (z_s)= \int_0^{z_s} \dd z_l\int_0^\infty \dd \sigma \Phi(\sigma,z_l)(1+z_l)^3
  \frac{c \dd t}{\dd z_l} \pi D_l^2 \theta_E^2(\sigma,z_l) \text{ ,}
\end{split}
\end{equation}
where $z_s$ and $z_l$ are respectively the source and lens redshifts, $\Phi(\sigma,z)$ is the velocity dispersion function.
Flux limited samples are also subject to magnification bias, which increases the relative probability that detected galaxies are gravitationally lensed, and concentrates sources in a flux limited sample around foreground objects (\citealt{Webster_1988}, \citealt{Wyithe_Nature_2011}). \\
\indent The observed luminosity function can therefore be approximated as the sum of the apparent luminosity function of the fraction of lensed galaxies and the intrinsic UVLF
\begin{equation}\label{eq:LF_apparent}
  \Psi_{obs}(L,R_{hl}, f) \approx \left(1-\tau_m l^2\right)\Psi\left(L\right)+\Psi_{lensed}(L,R_{hl}, f) \text{ .}
\end{equation}
We ignored the effective small demagnification in the unlensed UVLF in the right-hand side due to the conservation of flux over the whole sky since its effect is negligible over the bright-end of the apparent UVLF.
We adopted the standard Schechter function (\citealt{Schechter_1976}) for the intrinsic luminosity function
\begin{equation}\label{eq:Schechter_LF}
  \Psi(L)\dd L = \Psi_\star \left(\frac{L}{L_\star}\right)^\alpha\exp\left(-\frac{L}{L_\star}\right)\frac{\dd L}{L_\star} \text{ .}
\end{equation}
The evolution of the Schechter function parameters with respect to redshift is taken from \cite{Bouwens_2022} (in particular, see Fig. 6 and Sect. 3.4 in that paper). \\
\indent The apparent luminosity function of the fraction of lensed galaxies for a fixed value of ellipticity $f$ of the lens population is then
\begin{equation}\label{eq:LF_lensed}
\begin{split}
  \Psi_{lensed}(L,R_{hl},f) &= \int_0^{z_s} \dd z_l\int_0^\infty \dd \sigma \int_0^\infty \dd \mu \Phi(\sigma,z_l)(1+z_l)^3
  \frac{c \dd t}{\dd z_l}\\ & \times \mathcal{A} D_l^2 \theta_E^2(\sigma,z_l)
    \frac{1}{\mu}\dv{P}{\mu}\left(R_{hl}, f, \sigma, z_l,z_s\right)\Psi\left({\frac{L}{\mu}}\right) \text{ ,}
\end{split}
\end{equation}
where $\mathcal{A} = \left({D_l}/{D_s}\times l\right)^2$ is the dimensionless source plane area used in the previous section to derive $\frac{\textit{d}P}{\textit{d}\mu}$. 
Note that this expression for the bias is different than the one in \cite{Wyithe_Nature_2011}, as we can't separate the optical depth term from the integral since in the case of finite sources the magnification probability distribution is a function of the geometry and mass of the lenses. \\
\indent We adopt the \cite{Mason_2015} velocity dispersion function and its evolution with redshift
\begin{equation}\label{eq:Vel_disp_function}
  \Phi(\sigma,z) = \frac{\ln(10)}{p}\frac{\Phi^\star_S(z)}{\sigma(1+z)^\beta} 
  \left(\frac{\sigma}{\sigma^\star}\right)^{p^{-1}(1+\alpha_S)} 
  \exp\left[-\left(\frac{\sigma}{\sigma^\star}\right)^{p^{-1}}\right] \text{ ,}
\end{equation}
with their best fit values for the parameters 
$p=0.24$, $\beta=0.20$, $\alpha_S = -0.54$, $\sigma^\star=216 \text{ km/s}$, and
$\Phi^\star_S(z)=3.75\times10^{-3}(1+z)^{-2.46}\text{Mpc}^{-3}$.
This is close to the local Velocity Dispersion Function measured by the SDSS (\citealt{Choi_2007_SDSS}). \\
\indent The relation between galaxy size and luminosity is commonly fitted by a power law
\begin{equation}\label{eq:Lum_size_relation}
  R_{hl} = R_0 \left(\frac{L}{L_0}\right)^{\gamma} \text{ ,}
\end{equation}
where $R_0$ is the effective radius at $L_0$, and $\gamma$ is the slope. 
Here $L_0$ is the characteristic UV luminosity for Lyman-break galaxies at $z=3$, which corresponds to magnitude $M_0 = -21$ (\citealt{1999Steidel}).
As a reference, we adopted the best-fit values of $R_0=0.61$ kpc and $\gamma=0.25$ at $z=7$ found in \cite{Liu_Wyithe_Dragons_VII_2017}.
In this work we favoured the use of this simple relation over more complex luminosity-size relations (e.g. \citealt{IllustrisTNG_gal_size}). \\
\indent The fraction of observed lensed galaxies per luminosity interval at the bright end is well approximated by the ratio $\Psi_{lensed}/\Psi_{obs}$, as shown in the top panel of Fig. \ref{fig:mag_bias_and_ratio}.
In the lower panel of Fig. \ref{fig:mag_bias_and_ratio} we show the ratio of LFs with respect to a point source case. We see that extended sources can suppress the effect of magnification bias by several orders of magnitude.

\subsection{Comparison with observed luminosity functions}

We next consider the lensed UVLF for a population of lenses with varying ellipticities.
Based on the ellipticity distribution for the deflector population, we do a weighted sum over $f$ to obtain the predicted observed UVLF
\begin{equation}\label{eq:Weighted_sum_over_f}
\Psi_{obs}(L,R_{hl}) = \int \dd f p(f) \Psi_{obs}(L,R_{hl},f) \text{ .}
\end{equation}
We included the dependency on the ellipticity distribution of lenses using the geometry constraints up to $z=2$ provided by the SDSS (\citealt{van_der_Wel_2014}) and assuming that the ellipticity of the lensing potential is approximated by the projected star distribution ellipticity \footnote{The assumption is reasonable considering a constant mass-to-light ratio and a weak contribution of the dark matter halo in the central region where strong lensing happens}. Within the distributions listed by \citealt{van_der_Wel_2014}, we choose the one associated with the stellar mass bin $10<M_\star<10.5$ at redshift $1.5<z<2$ as it represents the galaxy population of the most effective lenses (due to their strength and cosmological distances geometry).
The final distribution $p(f)$ is approximately a Gaussian centered on $\langle f \rangle =0.5$ with a standard deviation of $\sigma_f = 0.3$.

\begin{figure}
  \centering
  \includegraphics[width=0.85\linewidth]{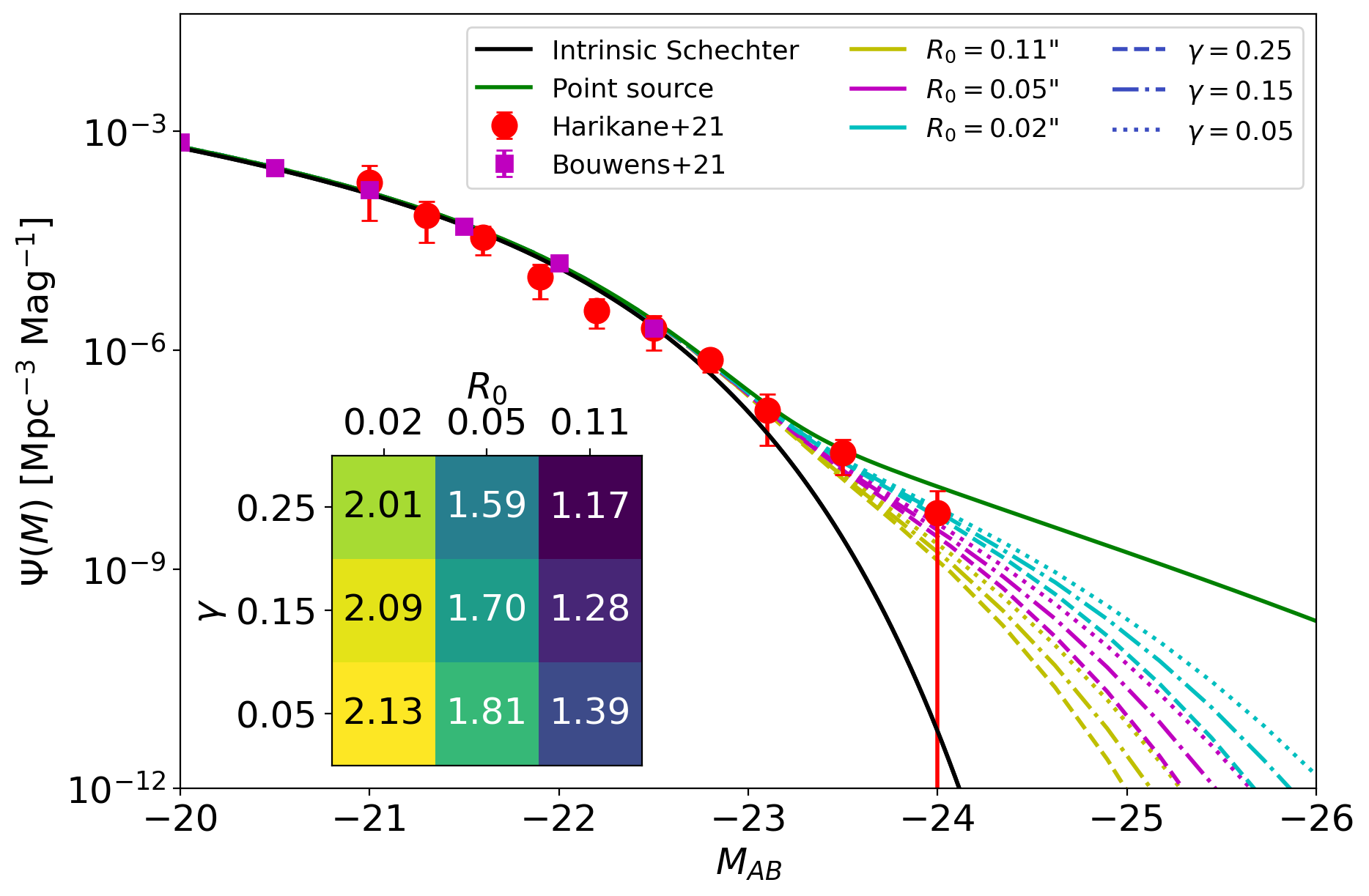}
    \caption{Lensed luminosity functions of extended sources assuming different choices of luminosity-size relation at redshift $z=6$.
    The intrinsic UVLF is shown in black, and it is chosen to be a Schechter function with parameters obtained in \citealt{Bouwens_2022}, as the best fit from the data from \citealt{Bouwens21_data} (magenta squares).
    The green line represents the apparent UVLF expected for a population of point-like sources.
    The yellow, magenta and cyan lines represent a luminosity-size relation with $R_{0}=0.11",0.05",0.02"$ respectively.
    The dashed, dashed-dotted and dotted lines represent a luminosity-size relation with $\gamma = 0.25, 0.15, 0.05$ respectively.
    For comparison, we show observational results for the UV luminosity function for galaxies obtained from dropout sources in \citealt{Harikane_2022} (red dots).
    The inset plot shows the value of the Bayes factor comparing the extended source model and the point source model (higher values corresponds to more likely models).}
    \label{fig:observation_z6}
\end{figure}

\begin{figure*}
  \includegraphics[width=0.78\linewidth]{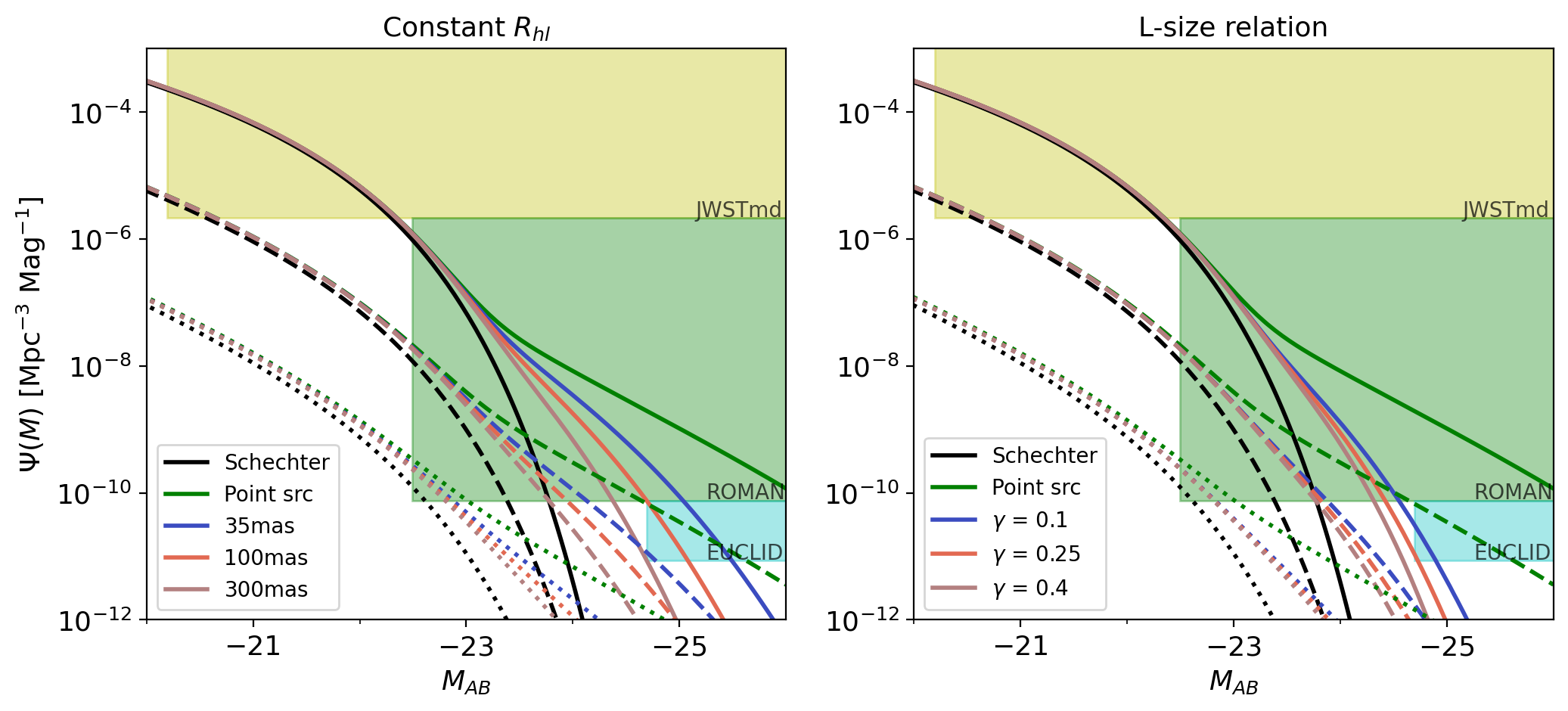}
    \caption{Predictions for the observed LFs caused by a foreground population of elliptical lenses for sources with constant half-light radius (left) and assuming a luminosity-size relation (right), for three values of redshift: $z=7$ (solid), $z=11$ (dashed) and $z=14$ (dotted). 
    The intrinsic UVLF is shown in black, and it is chosen to be a Schechter function with parameters obtained in \citealt{Bouwens_2022}. 
    The green line represents the apparent UVLF expected for a population of point-like sources.
    The blue, red, and brown lines represent a source with $R_{hl}=35,100,300$ (left) or alternatively with different choices for the value of $\gamma$ in the L-size relation $\gamma = 0.1, 0.25, 0.4$, with $R_0=0.11"$.
    We show the regions of magnitude and volume observable by future surveys, calculated for $z=7$:
    the order of magnitude of a typical medium-deep JWST Cycle 1 field (in this case \citealt{Windhorst_2022_PEARLS}, yellow), the High Latitude Wide Area Survey for the \textit{Roman Space Telescope} (green), and the Wide Survey from \textit{Euclid} (cyan).
    Since the survey depths shown here are calculated around $z=7$, the \textit{Roman Space Telescope} will see the bright end only for for $z\lesssim9$. 
    Euclid Wide Survey on the other hand, is not deep enough in magnitude to see the bright end, assuming galaxies are extended sources with some L-size relation.}
    \label{fig:LF_vs_surveys}
\end{figure*}

In Fig. \ref{fig:observation_z6} we compare our prediction for the galaxy UVLF at $z=6$ to observational results at that redshift from \cite{Harikane_2022} and \cite{Bouwens21_data}. 
The inset plot shows the value of the Bayes factor comparing the extended source model and the point source model. 
The effect of extended sources is to suppress the apparent brightening of the UVLF.
This is due to the cutoff in the magnification probability distribution shown in Fig. \ref{fig:dPdMu_ellipticity}, which convolved with the exponential shape of the Schechter profile gives the deviation from the power-law bright end expected from point-like sources.
This provides a constraint on the intrinsic size of high-z galaxies because of the small number of observations of bright ($M_{AB}\lesssim M_{AB,\star}-2$) sources.
The brightest data-point is compatible with a non-detection, therefore we do not penalize the models that predict a lower density at that magnitude, but the second brightest point marginally favors galaxies that are smaller compared to the predicted L-size relation parameters values from observed high-z galaxies (\citealt{2018Kawamata}, \citealt{Yang_GLASS_Lsize}, or the clumpy and extended morphologies found in \citealt{Bowler_clumpy_size_z7_2017}), simulations (\citealt{Liu_Wyithe_Dragons_VII_2017}, \citealt{Marshall_Meraxes}, or the dust-attenuated $L_{UV}$-size relation in \citealt{Marshall_Blue_Tides}).
That is, for a galaxy with $M_{AB}=-21$ at $z=6$, we find a size of $\approx 115$ pc instead of $\approx 800$ kpc found in \cite{Liu_Wyithe_Dragons_VII_2017} and \cite{2018Kawamata}. \cite{Yang_GLASS_Lsize} reports a size of around $\approx 400$ kpc in rest frame optical, and $\approx 200$ pc in rest frame UV, using a sample of galaxies with $z>7$ from early JWST data.
The lensed Schechter presented in \cite{Harikane_2022} (following \citealt{Wyithe_Nature_2011}) is lower than the one predicted in our work, due to the difference in the adopted magnification probability distribution.\\
\indent In the future we expect the UVLF at the bright end will probe the source size.
To illustrate, in Fig. \ref{fig:LF_vs_surveys} we show the observed LFs at $7<z<14$ for both a constant source size and a size-luminosity relation, along with a comparison to some current and future surveys aiming to probe the first galaxies volume density at the bright end of the UVLF. 
We see that the LFs from future wide field high-z surveys could be strongly affected by gravitational lensing.\\
\indent We note that observations of the bright end of the galaxy luminosity function may suffer AGN contamination (\citealt{Harikane_2022}; see also \citealt{Leethochawalit_2022} and \citealt{Bagley_2022} for potential evidence of such contamination at redshift $z=8$ and $z=9$).
Additional studies of the intrinsic AGNs luminosity function are needed to correctly disentangle the two population. It is worth remarking that AGNs would behave as point-like sources. However lensing bias would play a negligible effect on them assuming an intrinsic double power-law luminosity function profile (e.g. see \citealt{Wyithe_Loeb_2003}, \citealt{Qin_Wyithe_Dragons_X_2017}, and \citealt{Ren_2021}).\\
\indent Furthermore, recent studies by \cite{Mason_Trenti_Treu_2022} and \cite{Ferrara_2022} predict that the bright end of the UV luminosity function of high-z galaxies is strongly affected by dust attenuation. Significant dust attenuation has also been observationally confirmed in bright high-redshift galaxies (e.g. \citealt{Bowler_dust_2022}).
Since this modification on the UVLF profile happens in the rest-frame of the source, it will introduce a steepening in the intrinsic UVLF bright end, thus increasing the effects of gravitational lensing. 
If on the other hand the slope of the galaxy size-luminosity relation depends on dust attenuation (\citealt{2018Wu}, \citealt{Marshall_Blue_Tides}), we expect to see a change in the observed bright-end slope at different observed rest-frame wavelengths.

\section{Conclusions}
In this \textit{Letter} we calculate the effects of source size of bright high-redshift galaxies and lens ellipticity on gravitational lensing statistics. We show that these significantly suppress magnification bias, producing an apparent bright end with a sharper decline than previously considered.
Comparing our model for the observed UVLF to $z=6$ data from \cite{Harikane_2022}, which includes a departure from a Schechter profile at the bright end, we show that the luminosity function can be used to set constraints on the galaxies intrinsic size. 
This limited sample favors galaxies that are smaller compared to other predicted luminosity-size relation parameters values (\citealt{2018Kawamata}, \citealt{Liu_Wyithe_Dragons_VII_2017}, \citealt{Marshall_Meraxes}, or the dust-attenuated $L_{UV}$-size relation in \citealt{Marshall_Blue_Tides}). 
However, increasing the sample size will be necessary to make a reliable measurement. \\
\indent In the future, the difference in lensed LFs due to finite source size effects is expected to be measurable by the \textit{Roman Space Telescope} High Latitude Wide Area Survey, and would allow to constrain the size of galaxies in the epoch of reionization. 

\section*{Acknowledgments}
The authors thank Charlotte Mason for useful suggestions and discussions.
This research was supported by the Australian Research Council Centre of Excellence
for All Sky Astrophysics in 3 Dimensions (ASTRO 3D), through project number CE170100013.

\section*{Data Availability}
The modeled data discussed in this \textit{Letter} will be shared upon reasonable request to the corresponding authors.

\bibliographystyle{mnras}
\bibliography{LF_lensed_BE}

% Don't change these lines
\bsp  % typesetting comment
\label{lastpage}
\end{document}